\documentclass[letterpaper, 10 pt, conference]{ieeeconf}  

\IEEEoverridecommandlockouts                              

\usepackage{cite}
\usepackage{amsmath,amssymb,amsfonts}
\usepackage{optidef}
\usepackage{mathtools}
\usepackage{algorithmic}
\usepackage{graphicx}
\usepackage{textcomp}
\usepackage{xcolor}
\usepackage{adjustbox}
\usepackage{xspace}
\usepackage{soul}
\usepackage{bm}
\usepackage{bbm}
\usepackage{subfigure}
\usepackage{dsfont}
\usepackage[disable]{todonotes}
\usepackage[ruled]{algorithm2e}

\usepackage{hyperref}

\def\BibTeX{{\rm B\kern-.05em{\sc i\kern-.025em b}\kern-.08em
    T\kern-.1667em\lower.7ex\hbox{E}\kern-.125emX}}

\newcommand{\ellone}{$\mathcal{L}_1$\xspace}
\newcommand{\ellonedrac}{$\mathcal{L}_1$-DRAC\xspace}
\newcommand{\apriori}{\emph{a priori}\xspace}

\newcommand{\ito}{It\^{o}\xspace}
\newcommand{\Xt}[1]{X_{#1}}
\newcommand{\Ut}[1]{U_{#1}}
\newcommand{\Wt}[1]{W_{#1}}
\newcommand{\ULt}[1]{U_{\mathcal{L}_1,#1}}

\newcommand{\Hmu}[1]{H_\mu\left(#1\right)}
\newcommand{\Hsigma}[1]{H_\sigma\left(#1\right)}
\newcommand{\Borel}[1]{\mathcal{B}\left(#1\right)}

\newcommand{\Xstart}[1]{X^\star_{#1}}
\newcommand{\Ustart}[1]{U^\star_{#1}}
\newcommand{\Wstart}[1]{W^\star_{#1}}

\newcommand{\norm}[1]{\left\| #1 \right \| }
\newcommand{\Frobenius}[1]{\left\| #1 \right \|_F }

\newcommand{\Xdist}[1]{\mathbb{X}_{#1}}
\newcommand{\Xstardist}[1]{\mathbb{X}^\star_{#1}}
\newcommand{\Probability}[1]{\mathbb{P}\left(#1\right)}

\newcommand{\ELaw}[2]{\mathbb{E}_{#1}\left[ #2 \right] }

\newcommand{\Boldomega}{\omega}
\newcommand{\expo}[1]{e^{#1}}
\newcommand{\Lhat}[1]{\hat{\Lambda}\left(#1\right)}
\newcommand{\BoldTs}{T_s}
\newcommand{\br}[1]{\left( #1 \right)}
\newcommand{\cbr}[1]{\left\{ #1 \right\}}

\newcommand{\indicator}[2]{\mathds{1}_{#1}\left( #2 \right)}

\newcommand{\Xhatt}[1]{\hat{X}_{#1}}
\newcommand{\Xtildet}[1]{\tilde{X}_{#1}}

\newcommand{\Measure}[1]{\mathcal{M} \left( #1 \right) }

\newcommand{\sfp}{\mathsf{p}}

\newcommand{\LpLaw}[3]{\norm{#3}_{L_{#1}}^{#2}}

\newcommand{\pWass}[3]{\mathbb{W}_{#1}\left(#2,#3\right)}

\newcommand{\arxiv}{\todo{ArXiV}}

\newtheorem{assumption}{Assumption}
\newtheorem{remark}{Remark}
\newtheorem{theorem}{Theorem}

\title{\LARGE \bf
Wasserstein Distributionally Robust Adaptive Covariance Steering
}

\author{Aditya Gahlawat$^{\dagger}$, Vivek Khatana$^{\dagger,\ddagger}$, Duo Wang$^{\dagger,\ddagger}$, Sambhu H. Karumanchi$^{\dagger}$,  
\\
Naira Hovakimyan$^{\dagger}$, Petros Voulgaris$^{\ddagger}$
\thanks{This work was supported in part by AFOSR Grant FA9550-21-1-0411, the National Aeronautics and Space Administration (NASA) under Grants 80NSSC22M0070 and 80NSSC20M0229, and by the National Science Foundation (NSF) under Grants CMMI 2135925 and  IIS 2331878.}
\thanks{$^{\dagger}$The authors are with the Department of Mechanical Science and Engineering, Grainger College of Engineering, University of Illinois at Urbana-Champaign, Urbana, IL-USA 
        {\tt\small \{gahlawat, vkhatana, duowang, shk9,  nhovakim\}@illinois.edu}}%
\thanks{$^{\ddagger}$The authors are with the Department of Mechanical Engineering, University of Nevada, Reno, NV-USA 
        {\tt\small \{vkhatana, duow,   pvoulgaris\}@unr.edu}}%
}

\begin{document}

\maketitle
\thispagestyle{empty}
\pagestyle{empty}

\begin{abstract}

We present a methodology for predictable and safe covariance steering control of uncertain nonlinear stochastic processes.
The systems under consideration are subject to general uncertainties, which include unbounded random disturbances (\emph{aleatoric} uncertainties) and incomplete model knowledge (state-dependent \emph{epistemic} uncertainties).
These general uncertainties lead to temporally evolving state distributions that are entirely unknown, can have arbitrary shapes, and may diverge unquantifiably from expected behaviors, leading to unpredictable and unsafe behaviors.  
Our method relies on an \ellone-adaptive control architecture that ensures robust control of uncertain stochastic processes while providing Wasserstein metric certificates in the space of probability measures.
We show how these distributional certificates can be incorporated into the high-level covariance control steering to guarantee safe control.
Unlike existing distributionally robust planning and control methodologies, our approach avoids  difficult-to-verify requirements like the availability of finite samples from the true underlying distribution or an \apriori knowledge of time-varying ambiguity sets to which the state distributions are assumed to belong.     

\end{abstract}


\section{INTRODUCTION}

From the inception of feedback control in the form of centrifugal governors to the contemporary highly complex black-box autonomy algorithms for robotics, the safe operation of dynamical systems is an omnipresent requirement for the adoption of any technology by the general society. 
Guaranteeing predictable and safe operation in an inherently uncertain world environment is challenging, further exacerbated by incomplete knowledge of the systems themselves. 
In response to such challenges, one usually appeals to methodologies that can ensure robustness against epistemic uncertainties (lack of knowledge but learnable) and aleatoric uncertainties (inherently random disturbances). 
However, even before the control synthesis, one must consider various representations of the uncertainties we desire robustness against.
Assuming the uncertainties lie in deterministic bounded subsets of the state space offers a relatively easier solution to the robust control synthesis problem.
However, attempting robust worst-case analysis assuming bounded disturbances and deterministic uncertainties often leads to overly conservative results, see e.g.~\cite{mitchell2005time, herbert2017fastrack, singh2019robust} and references therein.
Instead, one can study average-case (or high-probability) stochastic safety guarantees to alleviate the conservativeness since then one analyzes the distributions and their associated statistical properties instead of purely their supports.
Using stochastic representations of uncertainties can lead to safety guarantees with possible reduced conservatism.
Furthermore, representing the stochastic systems' states using the associated distributions (probability measures) allows the use of statistical learning theory~\cite{vapnik1999overview} to incorporate data-driven learned components~\cite{mohri2018foundations}.

\begin{figure*}
    \centering
    \subfigure[]{\includegraphics[width=0.48\textwidth]{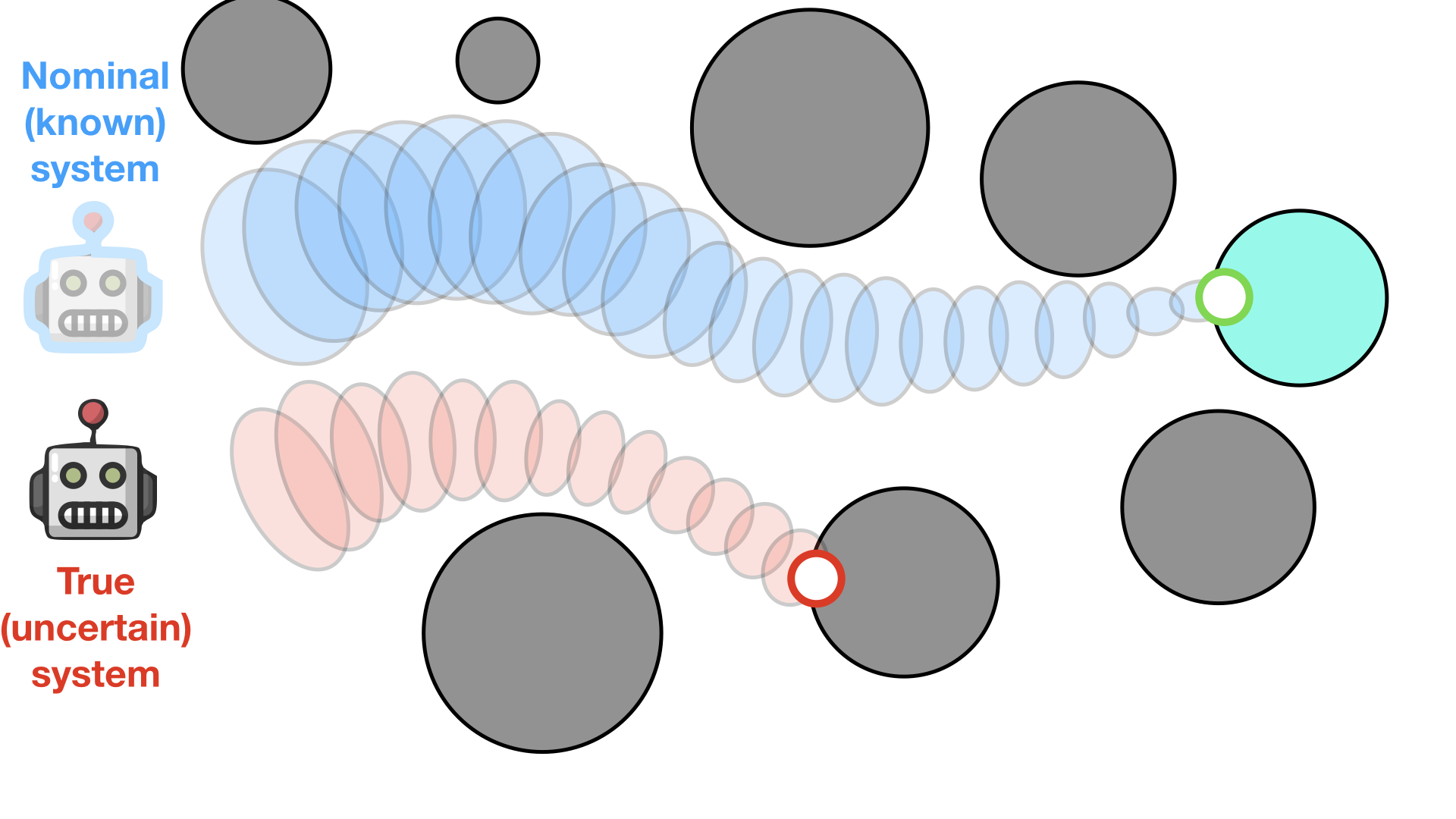}} 
    \subfigure[]{\includegraphics[width=0.48\textwidth]{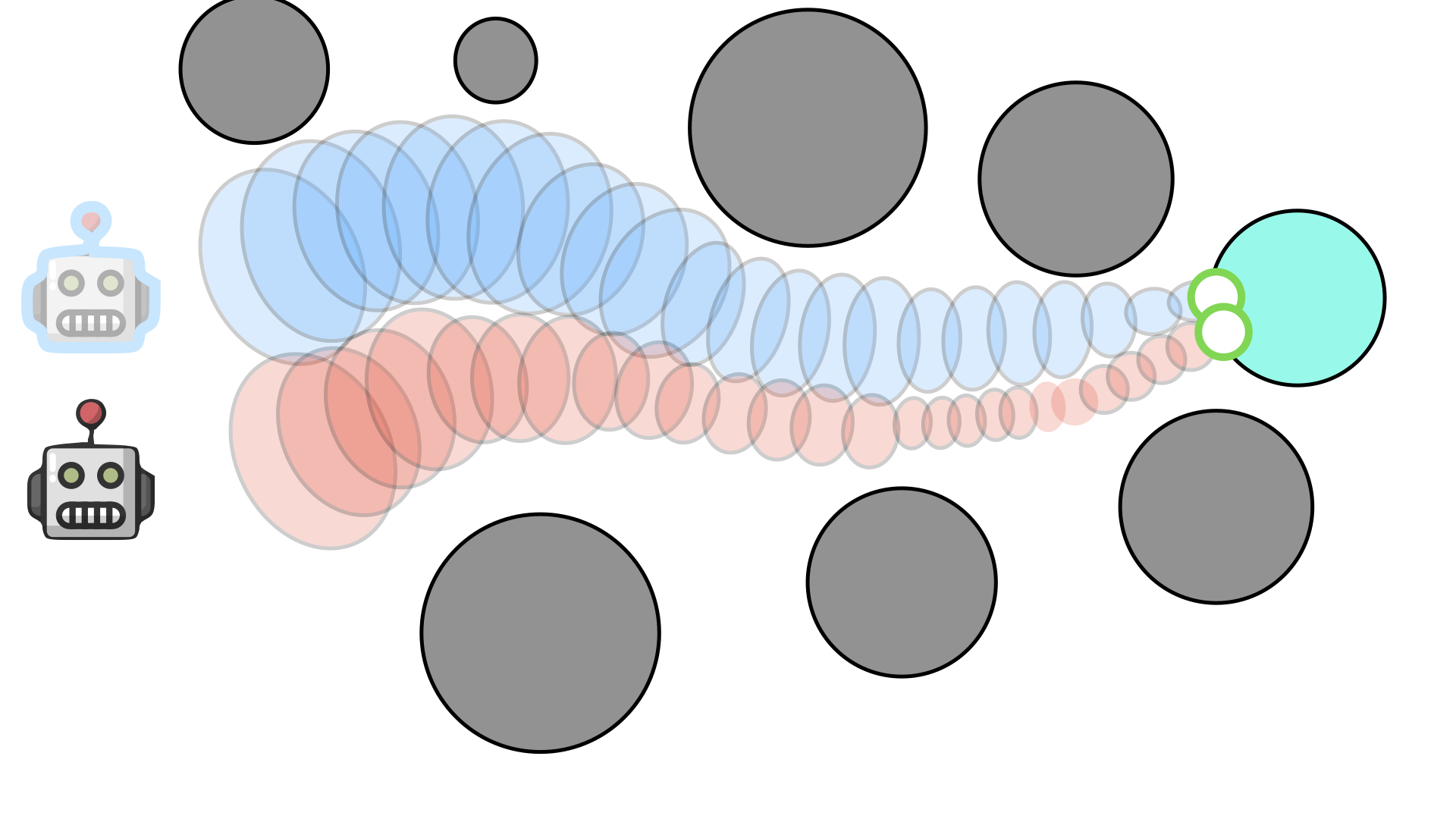}} 
    \caption{Consider the problem of safely navigating an uncertain system to a goal set (green circle), avoiding unsafe subsets (gray circles). One constructs a control law/policy for the uncertainty-free nominal (known) system as it is the best knowledge available for the true (uncertain) system. (a) While the control policy successfully guides the nominal (known) system to the goal set, illustrated with \ul{temporal state distributions} in light blue, applying the same policy to the true (uncertain) system leads to unquantifiable and undesirable deviation due to the presence of uncertainties (illustrated with light-red temporal state-distributions).  (b) Thus, it is critical to design an augmentation feedback policy to handle uncertainties such that the original policy can still guide the true (uncertain) system predictably and safely.}
    \label{fig:foobar}
\end{figure*}

Control of stochastic systems, stochastic optimal control (SOC) in particular, is a rich field with early examples such as the celebrated linear quadratic Gaussian (LQG)~\cite{kwakernaak1972linear, stengel1994optimal}, to the modern approaches like model predictive path integral (MPPI) control~\cite{williams2017model}, covariance steering~\cite{bakolas2016optimal,okamoto2018optimal,balci2021covariance}, and even reinforcement learning (RL)~\cite{levine2018reinforcement,bertsekas2019reinforcement}, since it can be argued that RL combines learning of models/reward functions and SOC synthesis. 
To ensure predictability and safety for stochastic systems, e.g., collision avoidance, a designer alters the SOC synthesis problem by incorporating binary evaluations (safe vs. unsafe) in the constraint set and continuous evaluations (e.g., the farther from an obstacle, the better) within the cost function~\cite{sun2015stochastic,pereira2021safe}. 
Given the stochastic nature of the systems, the notions of safety can be represented using \emph{risk metrics} like chance constraints (probability or moments of state $\in$ safe set)~\cite{vitus2015stochastic}, value at risk (VaR), and conditional value at risk (CVaR)~\cite{samuelson2018safety}. 
An in-depth investigation into risk metrics can be found in~\cite{MajumDarRisk}. 

Evaluating risk metrics requires the knowledge of the state distribution (probability measure induced by the random state).
Except for overly simple systems like linear plants initialized by Gaussian random variables, analytical descriptions of the time-varying state distributions are seldom available.
Hence, one usually resorts to approximating the distributions numerically, which is computationally expensive since this approach requires multiple samples from the true time-varying distributions.
A well-known example of approximating distributions via samples is sample average approximation~\cite{kleywegt2002sample}.
An approach to circumvent the limitation is to assume that the uncertainties and disturbances induce distributions with compact supports, thus allowing for a deterministic worst-case robust synthesis, see e.g.~\cite{mitchell2005time, herbert2017fastrack, singh2019robust}.
As we discussed, this approach leads to overly conservative results and ignores the statistical information in the stochastic representation. 

Distributionally robust optimization (DRO) allows one to take an alternate approach to robustness by directly hedging against the uncertainty due to the mismatch between the available \emph{nominal} (estimated/approximated) and the true distributions~\cite{delage2010distributionally,gao2023distributionally, rahimian2022frameworks}.
Since, in this case, the uncertainties are quantified by some notion of `distance' between distributions (probability measures), one considers an \emph{ambiguity set} - a set of probability measures, usually chosen as a ball centered around the nominal distribution. 
The radius of the ambiguity set is chosen such that the true distribution belongs to the ambiguity set with a desired probability. 
The use of DRO within SOC problems is usually referred to as distributionally robust control (DRC) in the literature. 
The area of DRC has seen considerable recent research effort since it allows one to cast the robust synthesis problem in the natural space of probability measures without requiring restrictive assumptions. See~\cite{shapiro2025distributionally,rahimian2022frameworks} for a few recent examples. 
One may define the ambiguity sets using various notions of `distances' between probability measures, e.g., moment-based ambiguity sets~\cite{nie2023distributionally}, ambiguity sets based on $f$-divergence~\cite{ben2013robust}, and Wasserstein metric-based ambiguity sets~\cite{zhao2018data,blanchet2019robust,yang2020wasserstein}. 
Moment-based ambiguity sets are often conservative and require a large sample size to produce reliable moment estimates. 
The use of $f$-divergence, in particular, KL-divergence, is popular to quantify the dissimilarity between probability distributions, but suffers from lack of symmetry in their arguments, the requirement of connected support of probability measures, and is not applicable between different types of probability measures, e.g., between discrete and continuous measures. 
Wasserstein distance-based approaches offer superior advantages since the Wasserstein distance defines a metric on the space of probability measures~\cite{villani2009optimal}. This allows one to quantify the dissimilarity between distributions in a rigorous fashion that accounts for the differences in the shapes of distributions as well~\cite{peyre2019computational}. 
Moreover, Wasserstein ambiguity sets contain a richer set of relevant distributions, and the corresponding Wasserstein DRO provides a superior sample performance guarantee~\cite{gao2023finite}.

While using Wasserstein DRO is attractive, establishing ambiguity sets containing the true distributions is challenging since the Wasserstein metric is the extremum of an infinite-dimensional optimization problem~\cite{gao2023finite}.
An approach to construct Wasserstein ambiguity sets is to leverage the finite sample guarantees to construct balls around empirical distributions such that the true distributions lie inside the ambiguity set with a high probability~\cite{mohajerin2018data}. 
The benefits of this approach stem from it depending purely on finite samples and has thus seen its application for a multitude of distributionally robust controller synthesis approaches~\cite{pilipovsky2024distributionally, chen2025distributionally,hakobyan2022wasserstein}. 
However, the assumption that samples are available from the true distributions can prove to be unverifiable for various applications of interest. 
For example, state distributions are time-varying by their nature of describing the evolution of dynamic processes.
Hence, requiring the availability of samples from the true distribution would translate to requiring multiple samples at each point in time, considering the system is described in a discrete-time fashion. 
Moreover, the number of samples required at each point corresponds directly to the size of the high-probability ambiguity sets via the finite sample guarantees~\cite{gao2023finite,van2015distributionally}. 
Another major hurdle that is relevant to the control of uncertain systems, as we consider in the manuscript, is that a control law will have access to either state or output measurements to compute the input to the system. 
Hence, the controller has access to precisely a \emph{single sample} from the true distribution at each point in time, rendering the Wasserstein finite-sample guarantees unusable.  

We provide \ul{Wasserstein distributionally robust adaptive covariance steering} - a methodology for distributionally robust control of uncertain nonlinear systems subject to both epistemic and aleatoric uncertainties. 
The features of our approach are as follows:
\begin{enumerate}
    \item We do not require the availability of any samples from the true distribution to construct the Wasserstein ambiguity sets;
    \item We instead rely on our recently developed \ellone-distributionally robust adaptive control (\ellonedrac)~\cite{L1DRAC}\arxiv   
    as a state-feedback control augmentation, that provides certificates of Wasserstein ambiguity sets with probability $1$. 
    The \ellonedrac controller is based on the architecture of the \ellone-adaptive control~\cite{L1book}.
    We have previously successfully applied \ellone-adaptive control to model-based reinforcement learning~\cite{sungrobust};
    \item The overall controller constitutes the coupling of the \ellonedrac controller with a high-level control law (covariance steering) in a feedback-motion planning approach~\cite[Chp.~8]{lavalle2006planning};
    \item The high-level covariance steering (CS) controller generates reference trajectories using only the \emph{nominal plant} knowledge, but incorporates Wasserstein DRO constraints using CVaR as enabled via the guaranteed ambiguity sets ensured by the low-level \ellonedrac;
    \item The proposed approach thus ensures safety of the system via the Wasserstein DRO while retaining the solvability of the CS controller;
    \item Since the high-level CS controller is altered only in a way to incorporate the additional Wasserstein DRO constraint, the proposed approach can be easily adapted to be used with most SOC controllers. 
\end{enumerate}

The manuscript is organized as follows: Section~\ref{sec:ProblemStatementPrelims} provides the problem statement that we aim to address, along with a brief discussion of the constituent covariance steering control and \ellonedrac. 
The main contribution is presented in Section~\ref{sec:DRACS}, which is then followed with the conclusions and exploratory directions in the final section.










\subsection{Notation}
We denote by $\mathbb{R}_{>0}$ and $\mathbb{R}_{\geq 0}$ the set of positive and non-negative reals, respectively. $\mathbb{N}_0$ is the set of natural numbers starting at $0$. $\mathcal{C}\left(\mathbb{R}^n;\mathbb{R}^m\right)$ denotes the set of continuously differentiable maps $\mathbb{R}^n \to \mathbb{R}^m$. We denote by $\Borel{F}$, the Borel $\sigma$-algebra generated by $F$ and the measures on $\Borel{F}$ by $\Measure{F}$. 
Furthermore, let $(S,\Sigma, \mathcal{M})$ be a measure space and $1 \leq p < \infty$, $\|f\|_{L_p}$ denotes the $L_p$ norm given by $\|f\|_{L_p} := \left(\int_{S} |f|^p d \mathcal{M} \right)^{1/p}$ and $\pWass{2\sfp}{\Xdist{t}}{\Xstardist{t}}$ denotes the $2$ Wasserstein metric between probability measures $\Xdist{t}$ and $\Xstardist{t}$.


\section{Problem Statement and Preliminaries}\label{sec:ProblemStatementPrelims}

Consider a complete probability space $\left(\Omega,\mathcal{F},\mathbb{P}\right)$, an adapted Brownian motion $\Wt{t} \in \mathbb{R}^{n_w}$, and a random variable $x_0 \sim \xi_0$ with finite second moment which is independent of $\sigma \left( \cup_{t\geq 0} \mathcal{W}_t \right)$, where $\mathcal{W}_t = \sigma \left( \Wt{s}~:~s\leq t  \right)$.  
We consider an uncertain system for which the evolution of dynamics is governed by the following nonlinear \ito process:
\begin{align}\label{eqn:ProStat:TrueSystem}
        d\Xt{t}
        =&
        F_\mu\left( \Xt{t},\Ut{t} \right)dt + F_\sigma\left( \Xt{t} \right)d\Wt{t}
        \notag 
        \\
        =& \left[A_\mu \Xt{t} + B\left(\Ut{t} + \Hmu{\Xt{t}} \right) \right]dt
        \notag 
        \\
        &+ \left[A_\sigma  + B \Hsigma{\Xt{t}}  \right]d\Wt{t}, \quad \Xt{0} = x_0, 
\end{align}
where $\Xt{t} \in \mathbb{R}^n$ is the system state, $\Ut{t} \in \mathbb{R}^m$ is the control input, and $B \in \mathbb{R}^{n \times m}$ is the input operator. 
The matrices $A_\mu \in \mathbb{R}^{n \times n}$ and $A_\sigma \in \mathbb{R}^{n \times n_w}$ represent the \emph{known} components of the dynamics, whereas, the functions $H_\mu \in \mathcal{C}\left(\mathbb{R}^n;\mathbb{R}^m\right)$ and $H_\sigma \in \mathcal{C}\left(\mathbb{R}^n;\mathbb{R}^{m \times n_w}\right)$ represent the \emph{uncertainties} in the drift and diffusion, respectively.
\begin{assumption}\label{assmp:uncertainties}
    \emph{
        The functions $H_\mu$ and $H_\sigma$ are globally Lipschitz continuous with known constants $L_\mu$ and $L_\sigma$, respectively. 
        Moreover, there exist known positive constants $\Delta_\mu$ and $\Delta_\sigma$ such that $\norm{\Hmu{a}}^2 \leq \Delta_\mu^2 \left(1 + \norm{a}^2 \right)$ and $\Frobenius{\Hsigma{a}}^2 \leq \Delta_\sigma^2 \left(1 + \norm{a}^2 \right)^\frac{1}{2}$, $\forall a \in \mathbb{R}^n$.
    }
\end{assumption}
\begin{remark}
    \emph{
        Under the assumption above, and the independence of the initial condition from the driving Brownian motion $\Wt{t}$, we assume that the input process $\Ut{t}$ is adapted to the underlying filtration and is regular enough so that $\Xt{t}$ exists as the unique strong solution of~\eqref{eqn:ProStat:TrueSystem}, see~\cite[Thm.~5.2.9]{karatzas1991brownian} and~\cite[Thm.~5.2.1]{oksendal2013stochastic}.
        Furthermore, see~\cite{L1DRAC} for the well-posedness under the \ellonedrac feedback.
    }
\end{remark}
For the unique strong solution $\Xt{t}$, we define its distribution as 
\begin{equation}\label{eqn:ProStat:Distribution}
    \Xt{t} \sim \Xdist{t}, \quad \Xdist{t} \in \Measure{\mathbb{R}^n}, \quad t \geq 0.
\end{equation}

Next, assuming there are $n_o \in \mathbb{N}$ obstacles to be avoided, we consider each of the obstacles defined via the union of $n_{l} \in \mathbb{N}$  half-spaces\footnote{We keep each obstacle's description uniform in $n_l$ (the number of half-spaces) for simplicity of exposition. The extension to a differing number of half-spaces for each obstacle is straightforward. }  which allows us to define the \emph{safe region} as follows:
\begin{subequations}\label{eqn:ProStat:SafeRegion}
    \begin{align}
        \mathbb{R}^{n} \supseteq \mathcal{X}_{safe} 
        \doteq
        \bigcup_{j=1}^{n_o}
        \mathcal{X}^j_{safe}, 
        \\
        \mathcal{X}^j_{safe}
        \doteq 
        \bigcap_{l = 1}^{n_l} 
        \left\{
            z \in \mathbb{R}^n~|~c_{j,l}^\top z \geq d_{j,l}
        \right\},
    \end{align}
\end{subequations}
where $c_{j,l} \in \mathbb{R}^n$ and $d_{j,l} \in \mathbb{R}$ are known \apriori.  
A representation of the safe region via a set of affine constraints is helpful in maintaining the numerical tractability of the high-level planner, and is thus common in the literature~\cite{okamoto2019optimal,hakobyan2020wasserstein}.
We further define the notion of safety via a user specified \emph{risk parameter} $\delta_s \in (0,1)$ by establishing the following requirement: 
\begin{equation}\label{eqn:ProStat:Safety}
    \text{Safe operation of~\eqref{eqn:ProStat:TrueSystem}} \Leftarrow \Probability{\Xt{t} \in \mathcal{X}_{safe}} \geq 1 - \delta_s,
\end{equation}
for all $t \in [0,T]$, where $T \in (0,\infty)$ is the horizon length.  
As stated previously, we will consider~\eqref{eqn:ProStat:Safety} with distributionally robust conditional value at risk (CVaR) that encodes safety via the Wasserstein metric.

\noindent \textbf{\emph{Problem Statement:}} 
Given Gaussian initial and terminal probability measures $\mathcal{N}\left(\mu_0,\Sigma_0\right)$ and $\mathcal{N}\left(\mu_T,\Sigma_T\right)$, respectively, for any $T \in (0,\infty)$,  the safety map $\Phi$, and the risk tolerance parameter $\delta_s$, compute the input process $\left(\Ut{t}\right)_{t \in [0,T]}$, such that the uncertain process $\Xt{t}$ in~\eqref{eqn:ProStat:TrueSystem} is optimal with respect to a user designed optimal control objective $J$, while satisfying the safety requirement~\eqref{eqn:ProStat:Safety}.  

We next provide a brief description of covariance steering control,  which is the chosen high-level planner, and the \ellonedrac control that enables the distributional robustness of the entire scheme.  

\subsection{Covariance Steering Control}\label{subsec:CovSteering}

Covariance steering (CS) control methodology is an approach to SOC that considers system state's mean and covariance  as optimization variables~\cite{bakolas2016optimal,yu2024optimal}.
Controlling the first two moments can be interpreted as effectuating the average behavior of the system and the \emph{shape} of the uncertainties.
Since CS is a model-based approach to SOC, it relies on the availability of a known (nominal) model of the system dynamics. 
We define the \emph{nominal (known)} system, which is the uncertainty-free version of~\eqref{eqn:ProStat:TrueSystem} by dropping the unknown $H_\mu$ and $H_\sigma$ to obtain
\begin{align}\label{eqn:ProStat:NominalSystem}
        d\Xstart{t} 
        =&
        \left[A_\mu \Xstart{t} + B \Ustart{t} \right] dt
        + A_\sigma  d\Wstart{t}, \quad \Xstart{0} = x_0^\star, 
\end{align}
where $\Wstart{t} \in \mathbb{R}^{n_w}$ is another Brownian motion defined on $\left(\Omega,\mathcal{F},\mathbb{P}\right)$, and is independent of $\Wt{t}$. Additionally, the initial condition $x^\star_0 \sim \xi^\star_0$ is a random variable with a finite second moment and is independent of the filtration induced by $\Wstart{t}$.
Similar to~\eqref{eqn:ProStat:Distribution}, we denote by $\Xstardist{t} \in \Measure{\mathbb{R}^n}$ the distribution of $\Xstart{t}$, i.e. $\Xstart{t} \sim \Xstardist{t}$.
By applying the Euler-Maruyama scheme~\cite{platen1999introduction} to~\eqref{eqn:ProStat:NominalSystem}, we obtain the following discrete-time version of the nominal (known) plant:
\begin{align}\label{eqn:ProStat:Discrete:NominalSystem}
    \Xstart{k+1} 
    =
    \Xstart{k} + \left[A_\mu \Xstart{k} + B \Ustart{k} \right]\Delta T + A_\sigma \Delta \Wstart{k},
\end{align}
with $\Xstart{0} = x_0^\star$, where $k \in \mathbb{N}_{0}\Delta T$, $\Delta T \subset \mathbb{R}_{\geq 0}$ is the interval of temporal discretization, and $\Delta \Wstart{k} = \Wstart{k+1} - \Wstart{k} \overset{i.i.d.}{\sim} \mathcal{N}\left(0_{n_w},\Delta T \mathbb{I}_{n_w}\right)$ are the increments of $\Wstart{t}$~\cite{oksendal2013stochastic} for each $k \in \mathbb{N}_{0}\Delta T$.
Following the setup in~\cite{okamoto2018optimal,goldshtein2017finite},  we re-write~\eqref{eqn:ProStat:Discrete:NominalSystem} as 
\begin{align}\label{eqn:ProStat:Discrete:NominalSystem:Unified}
    X^\star_{k'}
    =
    \mathcal{A}_\mu\Xstart{0} + \hat{B}\mathcal{U}^\star_T + \mathcal{A}_\sigma \mathcal{W}^\star_T,
\end{align}
where $\mathcal{U}^\star_T \doteq \left[{U^\star_0}^\top, {U^\star_1}^\top,\dots,{U^\star_{k'-1}}^\top \right]^\top$, $\mathcal{W}^\star_T \doteq \left[ {W^\star_0}^\top, {W^\star_1}^\top,\dots,{W^\star_{k'-1}}^\top \right]^\top$, and matrices $\mathcal{A}_\mu$, $\hat{B}$, and $\mathcal{A}_\sigma$ are defined accordingly.
The variable $k' \in \mathbb{N}$ defines the horizon as $T = k'\Delta T$.  
For chosen positive definite $\mathcal{Q}$ and $\mathcal{R}$ of appropriate dimensions, consider the following optimal control problem:
\begin{subequations}\label{eqn:CS:cost}
    \begin{align}
        \bm{\mathcal{U}^\star_T} := \bm{U^\star_{0 \dots k'-1}} 
        = {\arg\min}_{\mathcal{X}^\star_T,\mathcal{U}^\star_T}
        \mathbb{E}\left[J\left(\mathcal{X}^\star_T,\mathcal{U}^\star_T\right)\right]
        \\
        J\left(\mathcal{X}^\star_T,\mathcal{U}^\star_T\right)
        :=
        \left(\mathcal{X}^\star_T\right)^\top \mathcal{Q} \mathcal{X}^\star_T
    + 
    \left(\mathcal{U}^\star_T\right)^\top \mathcal{R} \mathcal{U}^\star_T, 
    \end{align}
\end{subequations}
subject to the constraints
\begin{subequations}\label{eqn:CS:constraints}
    \begin{align}
        \Xstart{k+1} 
        =
        \Xstart{k} + \left[A_\mu \Xstart{k} + B \Ustart{k} \right]\Delta T + A_\sigma \Delta \Wstart{k},
        \\
        \ELaw{}{X^\star_0} = \mu_0, 
        \quad 
        \ELaw{}{X^\star_T} = \mu_T,
        \\
        \ELaw{}{X^\star_0 \left(X^\star_0\right)^\top} - \ELaw{}{X^\star_0}\ELaw{}{X^\star_0}^\top = \Sigma_0, 
        \\
        \ELaw{}{X^\star_T \left(X^\star_T\right)^\top} - \ELaw{}{X^\star_T}\ELaw{}{X^\star_T}^\top = \Sigma_T. 
    \end{align}
\end{subequations}
We use covariance steering (CS) as the choice for solving the optimal control problem stated above, see e.g.~\cite{okamoto2019optimal} and~\cite{goldshtein2017finite} for further details. 
While mostly used for linear dynamics, CS can also be applied to nonlinear plants as in~\cite{ratheesh2024operator}.
Note that we have not included the safety constraints in the CS problem~\eqref{eqn:CS:cost}-~\eqref{eqn:CS:constraints}, even though the aforementioned references explicitly consider the risk measures.
The reason, as we discussed in the introduction, is that safety considerations based on the CS problem will not be valid for the uncertain (true) system since the state distributions \textcolor{red}{of each are} dissimilar owing to the presence of uncertainties.
Thus, we wish to formulate constraints that satisfy the distributionally robust notions of safety.
For this purpose, we need a method to obtain the Wasserstein ambiguity sets, which in our approach, are provided by the \ellonedrac control,  discussed next.

\subsection{\ellone-Distributionally Robust Adaptive Control (\ellonedrac)}

Recall that the nominal (known) system in~\eqref{eqn:ProStat:NominalSystem} is the uncertainty free version of the uncertain (true) system in~\eqref{eqn:ProStat:TrueSystem}.
The input $\Ustart{t}$ is designed such that the nominal state $\Xstart{t}$ satisfies the desired control objectives, e.g., optimality while remaining safe via CS control discussed above. 
Due to the presence of the state-dependent nonlinear uncertainties $H_{\{\mu,\sigma\}}$ in the true system, along with their interaction with the driving Brownian motion $\Wt{t}$, the behavior of the true state $\Xt{t}$ under the input $\Ustart{t}$ will be unquantifiably different to that of $\Xstart{t}$. 
In order to quantify and control the divergence between the nominal and true system states, we consider an input of the form 
\begin{equation}\label{eqn:L1DRAC:Input}
    \Ut{t} = \Ustart{t} + \ULt{t},
\end{equation}
where $\ULt{t} \in \mathbb{R}^m$ is the \ellonedrac input and is defined as the output of the following \emph{low pass filter}:
\begin{align}\label{eqn:True:Filter}
    \ULt{t} 
    =  
    - \Boldomega \int_0^t \expo{-\Boldomega(t-\nu)} \Lhat{\nu} d\nu,  
\end{align}
where $\Boldomega \in \mathbb{R}_{>0}$ is the \textbf{filter bandwidth}.
The \emph{adaptive estimate} $\hat{\Lambda} \in \mathbb{R}^m$ is obtained via the following \emph{adaptation law}:
\begin{multline}\label{eqn:True:AdaptationLaw}
    \Lhat{t} 
    =  
    0_m \indicator{[0,\BoldTs)}{t} 
    \\
    +
    \lambda_s \br{1 - e^{\lambda_s \BoldTs}}^{-1}
    \sum_{i=1}^{\lfloor \frac{t}{\BoldTs} \rfloor}    
    \Xtildet{i\BoldTs}
    \indicator{[i\BoldTs,(i+1)\BoldTs)}{t},
\end{multline} 
where $\Xtildet{i\BoldTs} = \Xhatt{i\BoldTs} - \Xt{i\BoldTs}$, and $\BoldTs \in \mathbb{R}_{>0}$ is the \textbf{sampling period}, $ \Theta_{ad}(t) = \begin{bmatrix}\mathbb{I}_m & 0_{m,n-m}  \end{bmatrix} \bar{B}^{-1} \in \mathbb{R}^{m \times n}$, with $\bar{B} \in \mathbb{R}^{n \times n}$  defined using $B$ as in~\cite{L1DRAC}\arxiv.
The parameter $\lambda_s \in \mathbb{R}_{>0}$ contributes to the solution $\Xhatt{t}$ of the \emph{process predictor} given by: 
\begin{multline}\label{eqn:True:ProcessPredictor}
    \Xhatt{t} 
    =
    x_0
    +
    \int_0^t 
        \left(-\lambda_s \mathbb{I}_n \Xtildet{\nu}+ f(\nu,\Xt{\nu}) \right) d\nu
    \\
    +
    \int_0^t 
        \left(g(\nu) \ULt{\nu} + \Lhat{\nu}\right) d\nu,
\end{multline} 
where $\Xtildet{t} = \Xhatt{t} - \Xt{t}$.
We collectively refer to $\cbr{\Boldomega,\BoldTs,\lambda_s}$ as the \textbf{control parameters} for the \ellonedrac input~\eqref{eqn:True:Filter}~-~\eqref{eqn:True:ProcessPredictor}.
Next, we place a regular assumption on the stability of the nominal diffusion free (deterministic) dynamics, i.e.,~\eqref{eqn:ProStat:NominalSystem} with $A_\sigma \equiv 0$.
\begin{assumption}
    There exist $P,Q \in \mathbb{S}^n_{\succ 0}$, and known scalars $\alpha_1,\alpha_2 \in \mathbb{R}_{>0}$ such that 
    \begin{align*}
        \alpha_1 \norm{x}^2 \leq x^\top P x \leq \alpha_2 \norm{x}^2, \quad \forall x \in \mathbb{R}^n, 
    \end{align*}
    and $A_\mu^\top P + P A_\mu = -Q$.
\end{assumption}
The assumption implies that the deterministic (diffusion free) nominal system is exponentially stable with $P \in \mathbb{S}^n_{\succ 0}$ defining the quadratic Lyapunov function certificate. 
Usually, a static feedback operator $K$ is constructed such that the closed loop $A_\mu+BK$ is Hurwitz. 
In such cases, we can simply consider $A_\mu \leftarrow A_\mu+BK$.
The following result from~\cite{L1DRAC} establishes the guarantees of the \ellonedrac controller.
\begin{theorem}[\cite{L1DRAC}]\label{thm:L1DRAC}
    \arxiv Suppose that the strong solution $\Xstart{t}$ of the nominal system~\eqref{eqn:ProStat:NominalSystem} under the input $\Ustart{t}$ satisfies $\LpLaw{2\sfp}{}{\Xstart{t}} \leq \Delta_\star$, $\forall t \in [0,T]$ for some known $\sfp \in \mathbb{N}_{\geq 1}$ and $\Delta_\star \in \mathbb{R}_{>0}$.
    Furthermore, for arbitrarily chosen scalars $\rho_a$ and $\epsilon$, define 
    $\rho_r = \sqrt{\frac{\alpha_2}{\alpha_1}}\LpLaw{2\sfp}{}{x_0- x^\star_0} + \Delta_{A_\sigma} + \epsilon$ and $\rho = \rho_r + \rho_a + \Delta_{A_\sigma}$, where $\Delta_{A_\sigma} \in \mathbb{R}_{>0}$ is a known constant that depends on $\Frobenius{A_\sigma}$.
    We can construct strictly positive functions $\zeta_1(\omega) \propto 1/\sqrt{\omega}$ and $\zeta_2(T_s) \propto \sqrt{T_s}$, and positive scalars $\beta_1$ and $\beta_2$ that depend on the continuity and growth bounds in Assumption~\ref{assmp:uncertainties}, and set $\omega,T_s \in \mathbb{R}_{>0} $ to satisfy $\beta_1 > \zeta_1(\Boldomega)$ and $\zeta_2 > \Theta_2(T_s)$. 
    Then, under the input~\eqref{eqn:L1DRAC:Input}, the distribution $\Xdist{t}$ of the uncertain state $\Xt{t}$ satisfies
    \begin{equation}\label{eqn:DRAC:WassersteinBounds}
        \pWass{2\sfp}{\Xdist{t}}{\Xstardist{t}} \leq \rho, \quad \forall t \in [0,T],
    \end{equation}
    where $\Xstardist{t}$ is the distribution of the nominal (known) system.
\end{theorem}


\section{Distributionally Robust \ellonedrac Covariance Steering}\label{sec:DRACS}

\begin{figure}[t]
\includegraphics[width=\columnwidth]{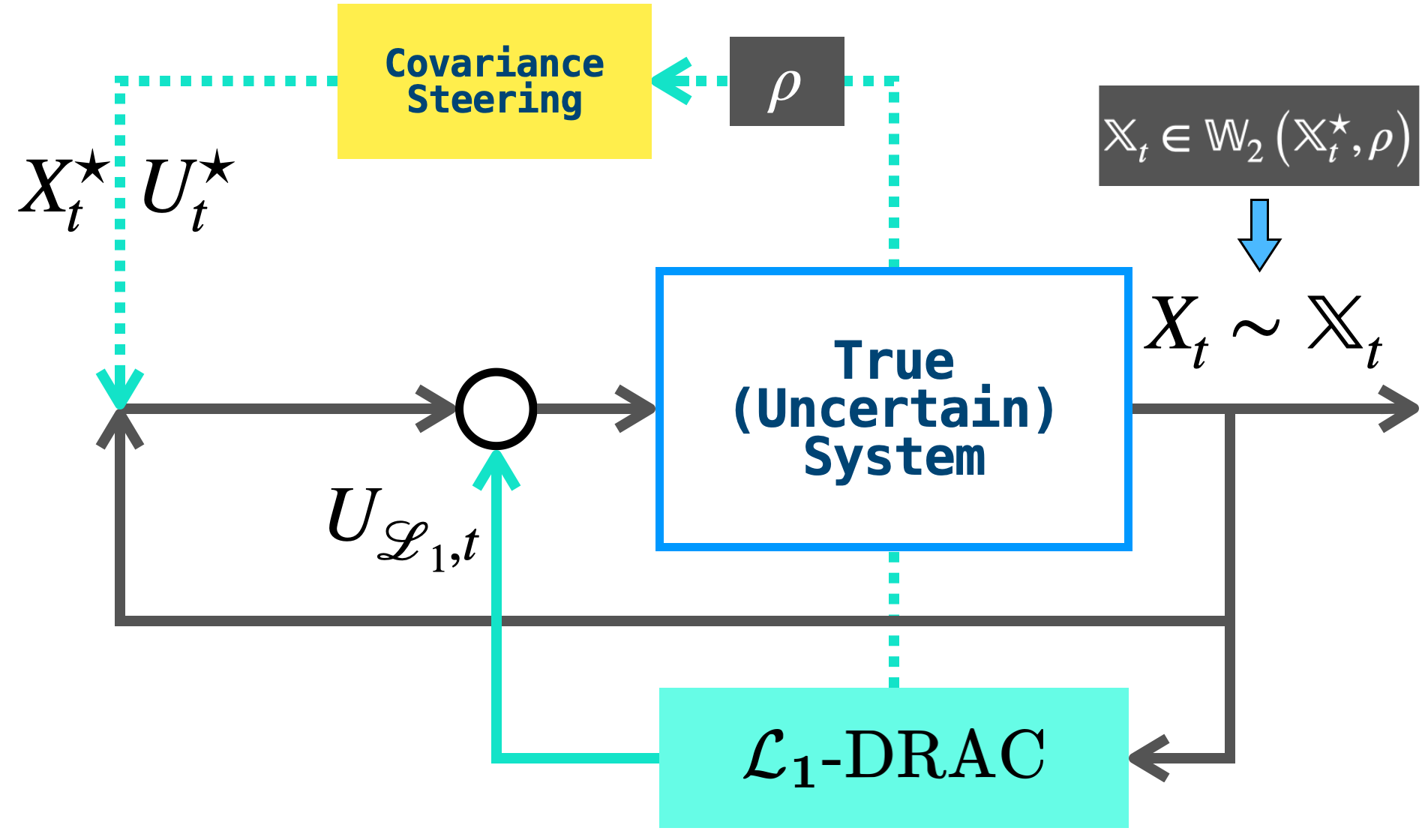}
\centering
\caption{ Architecture of the distributionally robust adaptive covariance steering methodology. The low-level \ellonedrac provides guarantees in terms of Wasserstein ambiguity sets of radius $\rho$ to the high-level covariance steering (CS) controller. The CS controller utilizes the ambiguity sets to generate nominal input robust to the distributional uncertainties by appending an additional constraint that does not sacrifice solvability of the entire scheme.}
\label{fig:arch}
\end{figure}

As we presented in the preceding section, the \ellonedrac controller ensures that the uncertain system state $\Xstart{t} \sim \Xstardist{t}$ remains \emph{uniformly bounded} with respect to  the nominal state $\Xstart{t} \sim \Xstardist{t}$ in terms of~\eqref{eqn:DRAC:WassersteinBounds}.
Thus, in order to ensure the safety of the uncertain system, we are now able to design inputs for the nominal system with the additional constraint imposed by the distributional bound~\eqref{eqn:DRAC:WassersteinBounds}, see Fig.~\ref{fig:arch}.
Recall from Sec.~\ref{subsec:CovSteering} that the covariance steering (CS) control algorithm is setup for discrete-time dynamics. 
Thus, in order to modify the CS control algorithm to incorporate the distributionally robust safety constraints, we once again use the Euler-Maruyama scheme~\cite{platen1999introduction} on the uncertain dynamics~\eqref{eqn:ProStat:TrueSystem} to obtain the following discrete-time dynamics
\begin{align}\label{eqn:ProStat:Discrete:TrueSystem}
    \Xt{k+1}
    =
    \Xt{k}
    +
    F_\mu\left( \Xt{k},\Ut{k} \right) \Delta T + F_\sigma\left( \Xt{k} \right)\Delta \Wt{k},
\end{align}
with $\Xt{0} = x_0$, where $k \in \left\{\mathbb{N}_{0}\right\}\Delta T$, $\Delta T \subset \mathbb{R}_{\geq 0}$ is the interval of temporal discretization and $\Delta \Wt{k} = \Wt{k+1} - \Wt{k}$ are i.i.d. $\mathcal{N}\left(0_{n_w},\Delta T \mathbb{I}_{n_w}\right)$ random variables for each $k \in \left\{\mathbb{N}_{0}\right\}\Delta T$.

It is important to note that the CS control algorithm applies to discrete-time dynamics, while the \ellonedrac control is designed for continuous-time stochastic systems. 
Thus, the bound in~\eqref{eqn:DRAC:WassersteinBounds} may not translate identically to its discrete counterparts.
Hence, one needs to account for the temporal-discretization errors.
Fortunately, the Euler-Maruyama scheme allows the explicit dependence of the strong error as a function of the temporal-discretization resolution $\Delta T$~\cite{kloeden1992stochastic}.
Then, to account for the discretization error, one needs to simply inflate the constant $\rho$ by the strong error.
We do not account for the strong error so as to not overly complicate the exposition, especially given the trivial nature of the solution. 

Now, let us define Wasserstein ambiguity sets of radius $\rho \in \mathbb{R}_{>0}$ centered around the nominal state probability measure $\Xstardist{k}  \in \Measure{\mathbb{R}^n} $:
\begin{equation}
    \mathbb{B}_\rho \left(\Xstardist{k} \right)
    =
    \left\{ 
        \nu \in \Measure{\mathbb{R}^n}~|~\pWass{2}{\nu}{\Xstardist{k}} \leq \rho.
    \right\}
\end{equation}
From~\eqref{eqn:DRAC:WassersteinBounds} in Theorem~\ref{thm:L1DRAC} with $\sfp=1$, we see that the input~\eqref{eqn:L1DRAC:Input} ensures that the state distribution $\Xdist{k}$ of~\eqref{eqn:ProStat:Discrete:TrueSystem} (or~\eqref{eqn:ProStat:TrueSystem} equivalently) satisfies $\Xdist{k} \in \mathbb{B}_\rho \left(\Xstardist{k} \right)$, $\forall k \in \cbr{1,\dots,k'}$.
For robust control of the uncertain system, we will thus use the guaranteed ambiguity set membership above to formulate the proposed distributionally robust covariance steering (CS) control.
We begin by first using the formulation in~\cite{okamoto2019optimal} to formulate the safety constraints.
Recall from~\eqref{eqn:ProStat:SafeRegion} that the safe region is a union of $n_o \in \mathbb{N}$ convex sets, each defined via the intersection of $n_l \in \mathbb{N}$ half-spaces.
For the horizon $T = k'\Delta T $ as in~\eqref{eqn:CS:cost}, let  
\begin{align*}
    \mathbb{R}^{n_o\times k'} \ni \mathcal{O}
    = 
    \begin{cases}
        \mathcal{O}_{j, k} = 1,\text{ if }\Probability{\Xstart{\cbr{k,k+1}} \notin \mathcal{X}_{safe}^{j}} \leq \delta_s, 
        \\
        \mathcal{O}_{j, k} = 0,\text{ otherwise }.
    \end{cases}
\end{align*}
It thus follows that the nominal system lies in a safe region with probability $1-\delta_s$, if the following constraint is satisfied~\cite{deits2015efficient, okamoto2019optimal}:
\begin{equation}
    \sum^{n_o}_{j = 1} \mathcal{O}_{j, k} = 1, \quad \forall k \in \cbr{1,\dots,k'}.
\end{equation}
To compute the probabilities for  the nominal system's location within a $j^{th}$-safe region, $j \in \cbr{1,\dots,n_o}$ we note that
\begin{multline}\label{eqn:ConvertChanceConstraint}
    \Probability{\Xstart{k} \notin \mathcal{X}_{safe}^j} 
    = \Probability{\bigcup^{n_l}_{l=1}\{c_{j,l}^\top \Xstart{k} > d_{j,l}\}}  
    \\
    \leq \sum^{n_l}_{l=1} \Probability{c_{j,l}^\top \Xstart{k} > d_{j,l}}.
\end{multline}

We encode the safety violation with respect to the $j^{th}$-safe safe region using conditional value at risk (CVaR)~\cite{nemirovski2007convex, xie2022optimized}, as follows:
\begin{equation}\label{eqn:CVaRConstraint}
    \mathrm{CVaR}^{X_k \sim \Xstardist{k} }_{1-\delta_s} \left[\max_{l \in \cbr{1,\dots,n_l}} (c^\top_{j,l}X_k - d_{j,l})\right] \leq 0,
\end{equation}
where $(1-\delta_s) \in \mathbb{R}_{>0}$ represents the tail threshold, and recall from~\eqref{eqn:ProStat:Safety} that $\delta_s \in (0,1)$ is the risk parameter. 
Using the CS approach in~\cite{okamoto2019optimal}, one obtains from~\eqref{eqn:ProStat:Discrete:NominalSystem} that $\Xstardist{k} \sim \mathcal{N}\left(\mu_k^\star, \Sigma_k^\star\right)$. 
The Gaussian nature of $\Xstardist{k}$ implies the following~\cite[Proposition~5]{norton2021calculating}:  
\begin{multline}\label{eqn:CVaRforGaussian}
    \mathrm{CVaR}^{X_k \sim \Xstardist{k}}_{1-\delta_s/n_l} 
    \left[ c^\top_{j,l}X_k - d_{j,l} \right] 
    \\
   = c^\top_{j,l}\mu_k^\star - d_{j,l} + \frac{\phi(\Phi^{-1}(1 - \delta_s/n_l))}{\delta_s/n_l}\sqrt{c^\top_{j,l}\Sigma_k^\star c_{j,l}},
\end{multline}
for $(j,l,k) \in \cbr{1,\dots,n_o} \times \cbr{1,\dots,n_l} \times \cbr{1,\dots,k'}$, where $\phi(\cdot)$ is the standard normal probability density function, $\Phi$ is the associated cumulative distribution function.
Note that  if the CVaR in~\eqref{eqn:CVaRforGaussian} with tail threshold $1-\delta_s/n_l$ holds $\forall (j,l)  \in \cbr{1,\dots,n_o} \times \cbr{1,\dots,n_l}$, then it upper bounds the CVaR in~\eqref{eqn:CVaRConstraint} as we have uniformly distributed the risk $\delta_s$ among the $n_l$ half-planes that define the $j^{th}$ safe region. 

Recall our previous discussion that~\eqref{eqn:DRAC:WassersteinBounds} in Theorem~\ref{thm:L1DRAC} with $\sfp=1$ leads to the conclusion that the input~\eqref{eqn:L1DRAC:Input} ensures that the state distribution $\Xdist{k}$ of~\eqref{eqn:ProStat:Discrete:TrueSystem} (or~\eqref{eqn:ProStat:TrueSystem} equivalently) satisfies $\Xdist{k} \in \mathbb{B}_\rho \left(\Xstardist{k} \right)$, $\forall k \in \cbr{1,\dots,k'}$. 
Thus, to ensure the constraint satisfaction for the uncertain system, we use the distributional guarantee of the \ellonedrac controller and the results on distributionally robust CVaR in~\cite{zhang2024short} to convert the nominal constraint in~\eqref{eqn:CVaRforGaussian} to the following distributionally robust constraint for the uncertain system:
\begin{align}\label{eqn:DRCVaRreformulated}
    \sup_{\nu \in \mathbb{B}_\rho \left(\Xstardist{k} \right)} &\mathrm{CVaR}^{X_k \sim \nu}_{1-\delta_s/n_l} (c^\top_{j,l}X_k - d_{j,l}) = c^\top_{j,l}\mu_k^\star - d_{j,l}  \nonumber \\
    &\hspace{-0.5in} +\frac{\phi(\Phi^{-1}(1 - \delta_s/n_l))}{\epsilon_{l,\text{fail}}}\sqrt{c^\top_{j,l}\Sigma_k^\star c_{j,l}} + \frac{\|c_{j,l}\|_2}{\sqrt{\delta_s/n_l}}\rho \leq 0, 
\end{align}
for $(j,l,k) \in \cbr{1,\dots,n_o} \times \cbr{1,\dots,n_l} \times \cbr{1,\dots,k'}$.
We can now incorporate the constraint above into the optimal control formulation in~\eqref{eqn:CS:cost}-\eqref{eqn:CS:constraints} to arrive at the following distributionally robust optimal control problem:
\begin{subequations}\label{eqn:finalCS:cost}
    \begin{align}
        \bm{\mathcal{U}^\star_T} := \bm{U^\star_{0 \dots k'-1}} 
        = {\arg\min}_{\mathcal{X}^\star_T,\mathcal{U}^\star_T}
        \ELaw{}{J\left(\mathcal{X}^\star_T,\mathcal{U}^\star_T\right)}
        \\
        J\left(\mathcal{X}^\star_T,\mathcal{U}^\star_T\right)
        :=
        \left(\mathcal{X}^\star_T\right)^\top \mathcal{Q} \mathcal{X}^\star_T
    + 
    \left(\mathcal{U}^\star_T\right)^\top \mathcal{R} \mathcal{U}^\star_T, 
    \end{align}
\end{subequations}
subject to the following constraints, for all $(j,l,k) \in \cbr{1,\dots,n_o} \times \cbr{1,\dots,n_l} \times \cbr{1,\dots,k'}$:
\begin{subequations}\label{eqn:finalCS:constraints}
    \begin{align}
        &\Xstart{k+1} 
        =
        \Xstart{k} + \left[A_\mu \Xstart{k} + B \Ustart{k} \right]\Delta T + A_\sigma \Delta \Wstart{k},
        \\
        &\ELaw{}{X^\star_0} = \mu_0,
        \quad 
        \ELaw{}{X^\star_T} = \mu_T,
        \\
        &\ELaw{}{X^\star_0 \left(X^\star_0\right)^\top} - \ELaw{}{X^\star_0}\ELaw{}{X^\star_0}^\top = \Sigma_0,  
        \\
        &\ELaw{}{X^\star_T \left(X^\star_T\right)^\top} - \ELaw{}{X^\star_T}\ELaw{}{X^\star_T}^\top = \Sigma_T,
        \\
        &\sum^{n_o}_{j = 1} \mathcal{O}_{j, k} = 1,
        \quad \mathcal{O}_{j,k} \in \{0,1\},
        \\
        &
        \begin{multlined}[b][0.85\linewidth]
            c^\top_{j,l}\mu_k^\star - d_{j,l} + \frac{\phi(\Phi^{-1}(1 - \delta_s/n_l))}{\delta_s/n_l}\sqrt{c^\top_{j,l}\Sigma_k^\star c_{j,l}}
            \\
            + \frac{\|c_{j,l}\|_2}{\delta_s/n_l}\rho 
            \leq 
            M(1 - \mathcal{O}_{j,k}), 
        \end{multlined}
        \\
        &
        \begin{multlined}[b][0.85\linewidth]
            c^\top_{j,l}\mu_{k+1}^\star - d_{j,l}
            \\
            + 
            \frac{\phi(\Phi^{-1}(1 - \delta_s/n_l))}{\delta_s/n_l}\sqrt{c^\top_{j,l}\Sigma_{k+1}^\star c_{j,l}}   
            \\ 
            + 
            \frac{\|c_{j,l}\|_2}{\sqrt{\delta_s/n_l}}\rho \leq M(1 - \mathcal{O}_{j,k}), 
        \end{multlined}
    \end{align}
\end{subequations}
where $M$ is a sufficiently large constant due to the Big-M formulation \cite{lofberg2012big}. 
As stated earlier, we use covariance steering as the choice for solving the optimal control problem stated above. In particular, we use the setup in~\cite{okamoto2019optimal} and solve the resulting CS optimization problem using numerical solvers \cite{dunning2017jump, mosekmosek,CoeyLubinVielma2020}. This yields control inputs of the form $\bm{\Ustart{t}} = \bm{U^\star_{t,\mu}} + \bm{U^\star_{t,\Sigma}}$, where, $\bm{U^\star_{t,\mu}}$ and $\bm{U^\star_{t,\Sigma}}$ steer the mean and covariance of the states $\Xstart{}$ respectively. A pseudo-code for the proposed methodology is provided in Algorithm~\ref{alg:DRACCS}.
\begin{algorithm}[h]
    \SetKwBlock{Input}{Input:}{}
    \SetKwBlock{Output}{Output:}{}
    \SetKwBlock{Initialize}{Initialize:}{}
    \SetKwBlock{Planning}{Plan For:}{}
    \SetKwBlock{Operation}{Operate With:}{}
    \SetKwBlock{STEPTWO}{STEP 2:}{}
    \Input{$\mbox{Horizon:} \ T$\\
    $\mbox{Initial and terminal conditions:} \ \mu_0, \Sigma_0, \mu_T, \Sigma_T, x^*_0$ \\
    $\mbox{Known dynamics:} \ A_\mu, A_\sigma, B$\\
    $\mbox{Safe set:} \ \mathcal{X}_{safe}$\\
    $\mbox{Risk tolerance parameters:} \ \delta_s$\\
    $\mbox{Discretization parameter:} \ \Delta T, k' := T/\Delta T$}
    \Planning{ $\bm{\mathcal{U}_T}^*$ \ by solving \ \eqref{eqn:finalCS:cost}-\eqref{eqn:finalCS:constraints} via CS \cite{okamoto2019optimal} }
    \Operation{ $U_t = \bm{\Ustart{t}} + \ULt{t}, \mbox{where} \ \ULt{t} \ \mbox{as in}~\eqref{eqn:True:Filter}$ \\
    $
    \mbox{and} \ \bm{\Ustart{t}} \ \mbox{is as defined in Theorem~\ref{thm:CVAR}}$. }
    \caption{Distributionally Robust \ellonedrac Covariance Steering}
        \label{alg:DRACCS}
\end{algorithm}
\begin{theorem}\label{thm:CVAR}
    Let $\bm{\mathcal{U}^\star_T} := \bm{U^\star_{0 \dots k'-1}} $ be the solution of the covariance steering optimal control problem~\eqref{eqn:finalCS:cost}-\eqref{eqn:finalCS:constraints}.
    Then, with the input $ \Ut{t} = \bm{\Ustart{t}} + \ULt{t}$, where $\bm{\Ustart{t}}$ is the piecewise constant signal generated by $\bm{U^\star_{0 \dots k'-1}}$, the state $\Xt{t}$ of the uncertain system~\eqref{eqn:ProStat:TrueSystem} (or~\eqref{eqn:ProStat:Discrete:TrueSystem} equivalently), satisfies the desired safety margin:
    \begin{equation}\label{eqn:thm:WassersteinCVaR}
        \mathrm{CVaR}_{1-\delta_s}^{\Xt{k} \sim \Xdist{k} }
        \left[X_k \in \mathcal{X}_{safe}\right] \leq 0, 
    \end{equation}
    for all $k \in \cbr{1,\dots,k'}$.
\end{theorem}
The proof is trivial by observing that Algorithm~\ref{alg:DRACCS} satisfies the constraint \begin{equation}\label{eqn:WassersteinCVaR}
    \max_{\nu \in \mathbb{B}_\rho \left(\Xstardist{k} \right)}
    \mathrm{CVaR}_{1-\delta_s}^{\Xt{k} \sim \nu }
    \left[ X_k \in \mathcal{X}_{safe} \right]\leq 0,
\end{equation}
which implies that~\eqref{eqn:thm:WassersteinCVaR} holds.
The equivalence of the two constraints is due to the inclusion $\Xdist{k} \in \mathbb{B}_\rho \left(\Xstardist{k} \right)$, $\forall k$, enforced by the \ellonedrac control evidenced by the uniform bound~\eqref{eqn:DRAC:WassersteinBounds}.


\section{Conclusion and Future Work}

We present a novel approach to Wasserstein distributionally robust safe control of uncertain stochastic systems. The key innovation lies in the use of our distributionally robust adaptive \ellonedrac controllers that enforce the inclusion of the uncertain system's distributions in uniform ambiguity sets around any nominal distributions generated by the high-level covariance steering (CS) controller. 
Thus, without the need for samples from the true distribution, and avoiding the need for empirical distributions, our approach can be implemented in a data-free regime while ensuring the satisfaction of safety constraints with probability $1$.

Our approach is amenable to multiple directions of further development; for example, the use of alternate high-level stochastic optimal controllers, robustness against the added uncertainty in the environment, and the use of learned components with the loop.






\bibliographystyle{IEEEtran}
\bibliography{References, references_DRAC}

\end{document}